\documentstyle[12pt]{article}
\def\be{\begin{equation}}
\def\ee{\end{equation}}
\def\ba{\begin{array}{c}}
\def\ea{\end{array}}
\def\p{\partial}
\def\r{\varrho}
\def\a{\varphi}
\def\ben{$$}
\def\een{$$}
\begin{document}

\titlepage
%\vspace*{4cm}

   \begin{center}
{\Large \bf
  Low-lying spectra

 in
 anharmonic three-body oscillators

 with a strong short-range repulsion
}

\end{center}

\vspace{5mm}

   \begin{center}

Miloslav Znojil

\vspace{3mm}

{\small \it \'{U}stav jadern\'e fyziky AV \v{C}R, 250 68 \v{R}e\v{z},
Czech Republic\\

e-mail: znojil@ujf.cas.cz}

\vspace{5mm}

\end{center}

 % {\today}, file Aje3r.tex

PACS {03.65.Ge}

\subsection*{Abstract}

One-dimensional three-body Schr\"{o}dinger equation is studied,
with binding mediated by the power-law two-body long-range
attraction $V^{(L)}(x) = \alpha^2x^L$ in superposition with a
short-range repulsion $V^{(-K)}(x) = \beta^2/x^K$ (plus, possibly,
further subdominant power-law components). Such an unsolvable and
non-separable generalization of Calogero model (where one had $L =
K = 2$) is shown to become separable and solvable at all integers
$K > 0$ and $L > 0$ in the limit of the large repulsion $\beta \gg
1$.

\newpage

 \section{Introduction}

Solutions of an $A-$particle Schr\"{o}dinger bound-state problem
 \be
 \hat{H} \Psi(\vec{x}_1,\vec{x}_2,\ldots,\vec{x}_A)
 = E\,\Psi(\vec{x}_1,\vec{x}_2,\ldots,\vec{x}_A)\,
 \label{SEA}
 \ee
are vital in atomic and nuclear physics as well as in quantum
chemistry. Using some realistic input Hamiltonian $ \hat{H} $ one
often calculates the low-lying spectrum of energies as a
characteristic output. Its construction is not easy in general.
Even the results of some very advanced (usually, variational)
numerical methods need not be reliable and may require an
independent verification, say, via an at least partially exactly
solvable, simplified $A-$body model (their concise review appeared
recently in \cite{Ullate}).

Of course, the most natural opportunity for the (partial) exact
solvability occurs in many-body Schr\"{o}dinger equation
(\ref{SEA}) with the first nontrivial choice of $A=3$. In the
single dimension with $x_j \in I\!\!R$ and for the spinless and
equal-mass particles, the solvability may play an independent
important role in phenomenological as well as purely mathematical
applications of the pertaining three-body models
 \be
 H \Psi(x_1,x_2,x_3) = E\,\Psi(x_1,x_2,x_3)\,
 \ \ \ \ \ \ \ \ \ \  H =
 -
 \sum_{i=1}^{3}\ \frac{\p^2}{\p{x_i}^2}+\hat{W}(x_1,x_2,x_3)\,,
 \label{SE}
 \ee
especially when their binding is mediated by the mere two-body
forces,
 \be
 \hat{W}(x_1,x_2,x_3) =
 V(x_1-x_2)+
 V(x_2-x_3)
 + V(x_3-x_1)\,.
  \label{tribal}
  \ee
In such a setting, in 1969 \cite{Calogero}, Francesco Calogero
discovered the exact solvability of the model which was based on
the use of the most elementary spiked harmonic-oscillator (SHO)
two-body forces
 \be
 V^{(SHO)}(x_i-x_j) = \omega^2\,(x_i-x_j)^2
 + \frac{\nu(\nu+1)}{(x_i-x_j)^{2}}\,.
 \label{Calogero}
 \ee
This choice was motivated not only by the strong immediate
phenomenological appeal of the SHO two-body potential itself but
also by the encouraging fact that at the vanishing $\nu=0$, the
simplified Schr\"{o}dinger equation (\ref{SE}) -- (\ref{Calogero})
is well known to remain separable. This observation opened the
path towards the closed $\nu \neq 0$ solution since in the polar
Jacobi coordinates, its explicit construction emerges immediately
after the reduction  of the singular part of the sum
(\ref{tribal}) of potentials (\ref{Calogero}) to an elementary
expression~\cite{Calogero}.

Our present paper may be perceived as a direct continuation of the
same $A=3$ project, paying attention just to a broader class of
potentials characterized by their more general power-law behaviour
in both the short- and long-range domains. These potentials may be
chosen in the similar two-term form
 \be
 V(x_i-x_j) = \alpha^2\,(x_i-x_j)^L
 + \beta^2\,(x_i-x_j)^{-K},
 \ \ \ \ \ \ \ L, K = 1, 2, \ldots\
 \label{ahospikforces}
 \ee
or in the more general polynomial form (\ref{anhCalogero}) of
Appendix A. They may mimic many popular confining forces, say, of
the Lennard-Jones type. They combine a suitable strongly repulsive
singular core in the origin \cite{Hall} with the standard
anharmonic-oscillator asympotics. One arrives, in this way, at the
most natural generalization of the Calogero's forces, indeed.

We shall analyze the latter three-body problem in its two explicit
partial differential forms (\ref{PDE}) and (\ref{PDEpol}) of
Appendix A. The main purpose of such a study  was, originally, a
search for all the possible remnants of the exact solvability
which is,  definitely, lost in its completeness at any $L > 2$
and/or $K > 2$. The project found an initial encouragement also in
our unexpected observation that the transition to $L = 4$ need not
necessarily mean the loss of separability of eq. (\ref{SE}) (see
section 2 and, in particular, subsection 2.1 below). In parallel,
we believed that even after a complete loss of separability, the
problem might remain {\em approximately} separable and/or
solvable.

In this direction, two main technical ingredients of our study of
eq. (\ref{SE}) consist in the use of the Jacobi coordinates at any
$K$ and $L$ (cf. Appendix A) and in a successful generalization of
the above-mentioned Calogero's trigonometric $K = 2$ identity to
all the integer exponents $K$ and $L$ (cf. Appendix B). An
ultimate incentive for the presentation of all these results came
from our {\it a posteriori} observation that the latter identities
prove amazingly elementary and compact.

In a way guided by a few examples revealing the simplifying role
of the growth of $\beta$ (and described in more detail in
subsections~2.2 and 2.3) we are going to employ the simplicity of
our auxiliary trigonometric identities of Appendix B for
conversion or re-arrangement of our partial differential (PD)
Schr\"{o}dinger eq. (\ref{SE}) into a separable problem [i.e.,
into a set of two ordinary differential (OD) equations] plus
corrections.

This idea in its application to our non-separable $A=3$ problem is
fully developed in section~3. We demonstrate there how the
practical relevance of the separability-violating corrections
decreases very quickly with an increase of the dominant repulsive
component in superpositions (\ref{ahospikforces}). We show how our
separable large$-\beta$ approximants may be constructed as
solvable and how they generate the reliable low-lying three-body
spectra. For an explicit illustration of the efficiency of such a
strategy, we first test it on the solvable $K=L=2$ case (i.e., on
the model of Calogero) in subsection~3.1.

In subsection 3.2 we move to the first ``unsolvable" example with
$L = 3$ and show how the comparatively compact character of the
trigonometric identities of Appendix B enable us to achieve a full
quantitative understanding of the competing mechanisms of
repulsion and confinement in our non-separable two-dimensional
effective potential wells with complicated shapes.

Our final illustrative example of subsection 3.3 employs $K = L =
3$ and represents, therefore, a full fledged modification of the
Calogero's model in both its short- and long-range interaction
parts. Of course, this modification remains non-separable and is
also exactly solvable only in the limit $\beta \to \infty$.

Section~4 is the summary of our results which re-emphasizes that
all our three-body spiked anharmonic oscillators share the
features demonstrated via particular examples. Thus, for all of
them the low-lying spectra become available in the physically very
appealing asymptotic domain characterized by the very strong
short-range repulsion in our two-body interactions.

\section{Elementary guide: quartic oscillator}

\subsection{Quartic Schr\"{o}dinger equation
and its ``forgotten" separability}

We need not recall the techiques presented in Appendices A and B
to see that the potential function (\ref{tribal}) remains
virtually trivial, in the cartesian coordinates $X$ and $Y$, for
all the three-body regular harmonic oscillators (RHO) with $L=2$
and $\beta=0$ in (\ref{ahospikforces}). It is well known that the
related PD Schr\"{o}dinger equation (\ref{PDE}) with
   \[
{U}^{(2)}(X,Y)=3\,{X}^{2}+3\,{Y}^{2} \,
 \]
degenerates to a pair of the ordinary and solvable confluent
hypergeometric differential equations. In polar coordinates, in
addition, this two-dimensional RHO potential term becomes
completely independent of the angular variable $\a$,
  \be
  {\Omega}^{(2)}(\r,\a)=
{U}^{(2)}[X(\r,\a),Y(\r,\a)]= 3\,\r  ^2\,.
 \ee
Still, to our great surprise we revealed that this example is not
unique. There exists another two-body potential which leaves our
PD Schr\"{o}dinger equation (\ref{PDEpol}) separable. Such an
assertion can be easily verified since in the context of Appendix
B.1 we may derive that
 \[
{U}^{(4)}(X,Y)=\frac{9}{2}\,\left ({X}^{2}+{Y}^{2}\right )^{2},
\]
i.e.,
 \[
  {\Omega}^{(4)}(\r,\a)=
{U}^{(4)}[X(\r,\a),Y(\r,\a)]= \frac{9}{2}\,\r ^4\,.
\]
We see that the quartic oscillator belongs to a very exceptional set
of three-body problems which are {\em rigorously} reducible to the
solutions of an ordinary differential equation. Such an observation
will definitely deserve a separate treatment, say, in the context of
perturbation theory.

In what follows we are going to skip this opportunity because the
separability of the quartic three body oscillator seems so
exceptional and un-paralleled by any other power-law model.
Indeed, in the light of the discussion in Appendix B we may be
certain that the spiked quartic anharmonic oscillator (SQAO)
two-body interaction
 \be
 V^{(SQAO)}(x_i-x_j) =\omega^2\, \left( x_i-x_j \right )^2
 + \lambda\,\, \left( x_i-x_j \right )^4+
 \frac{\nu(\nu+1)}{ \left( x_i-x_j \right )^2}  \,
  \label{SQO}
 \ee
generates the only set of the PD Schr\"{o}dinger equations
(\ref{PDEpol}) which remain separable in the polar Jacobi
coordinates.

\subsection{The role of the pronounced two-body repulsion }

In a marginal remark let us add that in contrast to the
above-mentioned Calogerian $K=L=2$ model, the separability of the
latter quartic model does not imply its exact solvability of
course. Although its two-body input force (\ref{SQO}) is merely
slightly more complicated than its Calogerian SHO predecessor
(\ref{Calogero}), the SQAO bound state problem must be solved in
terms of perturbation expansions or by a numerical, approximate
method \cite{Fluegge}.

In spite of the (expected) loss of the exact solvability, the
spiked quartic illustration may still profit from the
above-mentioned experience gained during the study of the
exceptional force (\ref{Calogero}).  One learns, in particular,
that the apparently complicated expression
  \ben
 {U}^{(SHO)}(X,Y)=
 {3}\,\omega^2(X^2+Y^2) +\frac{\nu(\nu+1)}{2\,X^{2}}
 +\frac{\nu(\nu+1)}{2\,(X-\sqrt{3}Y)^{2}}
+\frac{\nu(\nu+1)}{2\,(X+\sqrt{3}Y)^{2}}\,
 \een
emerging in the Calogero's paper \cite{Calogero} may be
significantly simplified by its further re-parametrization [cf.
eq. (\ref{polar}) in Appendix A]. As long as we work now with the
same repulsive core, the same function occurs in SQAO case. We
have to return to the Calogero's ``experimentally" discovered
trigonometric identity
  \ben
 \frac{1}{2X^{2}}
 +\frac{1}{2(X-\sqrt{3}Y)^{2}}
+\frac{1}{2(X+\sqrt{3}Y)^{2}} \equiv
 \een
  \be
 \equiv
 \frac{1}{2\varrho^2} \,
 \left ( \frac{1}{\sin^2\a}
+\frac{1}{\sin^2(\a+\frac{2}{3}\pi)}
+\frac{1}{\sin^2(\a-\frac{2}{3}\pi)} \right )=
 \frac{9}{2\,\r^2\,\sin^2 3\,\a}\,
 \label{singular}
 \ee
which makes, obviously, also the SQAO version of the PD equation
(\ref{PDE}) manifestly separable.

Our fascination by the Calogero's model concerns, first of all, an
ease with which one may replace the difficult, PD Schr\"{o}dinger
equation (\ref{SE}) by a system of independent OD equations. In
this language, the introduction of the spherical Jacobi
coordinates replaces both the SHO and SQAO versions of the PD
problem (\ref{SE}) by the much more easily tractable ``radial" OD
equation
 \be
 \left [
 -\frac{d^2}{dr^2}+
 \frac{\nu(\nu+1)}{r^2} + \omega^2r^2 + \lambda\,r^4
 -E
 \right ]\,\psi(r) = 0\,\ \ \ \ \ r \in (0,\infty)
 \label{repres}
 \ee
with $\lambda=0$ or $\lambda > 0$, respectively.

It is well known that we have here $\nu = (A-3)/2+\ell$ where the
integer index $\ell=0,1,\ldots$ numbers the so called
(hyper)spherical harmonics \cite{Sotona}. One also recollects,
vaguely, that an addition of a new force $V_{spike}(r)\sim g/r^2$
would remain tractable in eq. (\ref{repres}) because its form
coincides precisely with the kinetic (so called centrifugal)
singular term so that we just have to admit any real $\nu=\nu(g)$
\cite{Hall}.

For the time being, let us fix $\lambda=0$ and pay attention just
to the exactly solvable spiked harmonic oscillators with a strong
repulsion $g\gg 1$ (i.e., with a large real $\nu \gg 1$). Of
course, their spectrum $E_n=\omega\,(4n+2\nu+3)$ with $n = 0, 1,
\ldots$ is equidistant and moves merely upwards with the steady
growth of $\nu$. Even without any knowledge of the exact formulae,
this phenomenon is easily explained by an elementary observation
that near the minimum, the {\em shape} of the SHO potential term
in eq. (\ref{repres}) may be very well approximated by a
one-dimensional RHO well. Thus, we may starts from the formal
Taylor series
 \ben
 V_{eff}(r)=
 \frac{\nu(\nu+1)}{r^2} + \omega^2r^2
 = V_{eff}(R)+ \frac{1}{2} (r-R)^2\,
 V''_{eff}(R) + {\cal O}\left [(r-R)^3\, V'''_{eff}(R)\right ]
 \een
 \be
= 2\omega^2R^2+4 \omega^2(r-R)^2 + {\cal O}\left
[\frac{(r-R)^3}{R}\right ]\,.
 \ee
{In the vicinity of the absolute minimum at $r=
R=[\nu(\nu+1)/\omega^2]^{1/4}$} and in the asymptotic domain
characterized by the ``thick" spike term, $\nu \gg 1$, the latter
potential is very well approximated by its first two terms since
$1/R \approx 1/\sqrt{\nu} \ll 1$.

One arrives at a consistent picture where the approximate energy
levels calculated for the {one-dimensional} leading-order RHO
well, i.e., the values
 \ben
 \tilde{E}_m = V_{eff}(R)+ 2\omega(2m+1), \ \ \ m = 0, 1, \ldots
 \een
{\em coincide} with the above exact low-lying SHO spectrum $E_n,
E_1, \ldots$ up to the second order term in $1/R$. This is an
extremely encouraging observation which shows that the domain
where the short-range repulsion is large is, in a way, privileged.

One should keep in mind that in our exactly solvable illustration,
the parameter $1/R$ is precisely the appropriate measure of
smallness of corrections. Once its value proves sufficiently
small, we arrive at an asymptotic equivalence of the two models,
 \be
 {\rm RHO}\ \
 \longleftrightarrow\ \
 {\rm SHO}\,.
 \label{doupleq}
 \ee
In addition, we may easily move beyond the leading-order precision,
making use of the textbook perturbation series \cite{Marcelo}. Their
explicit construction to arbitrary order and for virtually any
potential may proceed, e.g., along the lines described in full detail
in the review paper~\cite{Bjerrum}.

Same observations may be easily extended to the ``purely
numerical" SQAO problem where the effective potential $V_{eff}(r)=
\frac{\nu(\nu+1)}{r^2} + \omega^2r^2 +\lambda\,r^4$ with $r = R +
x$ may equally well be expanded in the similar Taylor series
 \ben
  \left [{\frac {\nu\,\left (\nu+1\right )}{{R}^{2}}}+{{
\omega}}^{2}{R}^{2}+{ \lambda}\,{R}^{4}\right ]+2\,\left [{{
\omega}}^{2}R-{\frac {\nu\,\left (\nu+ 1\right
)}{{R}^{3}}}+2\,{\lambda}\,{R}^{3}\right ]x + \ldots\,.
 \een
We see that the point $R$ becomes an absolute minimum of
$V_{eff}(r)$ whenever the linear, ${\cal O}(x)$ term vanishes.
This specifies the value of $R$ exactly (i.e., as a positive root
of the cubic algebraic equation) and enables us to find the
subsequent quadratic and higher Taylor-series terms
 \be
  \ldots
+\left [{{\omega}}^{ 2}+3\,{\frac {\nu\,\left (\nu+1\right
)}{{R}^{4}}}+6\,{\lambda}\,{R}^{2 }\right ]{x}^{2}+4\,\left
[-{\frac {\nu\,\left (\nu+1\right )}{{R}^{5} }}+{\lambda}\,R\right
]{x}^{3}+O\left ({x}^{4}\right )
 \ee
in closed form.

\subsection{Another example possessing an inverse-cubic core}

In previous section we have chosen the exactly solvable
illustration just to emphasize the key idea of construction which
lies in the coincidence of the shapes of two different potentials
near their minima, {\em caused} by the growing strength of the
singularity in the origin. Let us now verify an applicability of
such an approach using a slightly less trivial form of the
single-particle repulsion,
  \be
 \left [
  -\frac{d^2}{dr^2} + W(r)- E
 \right ] \,\Psi(r) = 0, \ \ \ \ \ \ \ \ \
 W(r)=  F\,r^{3}+\frac{G}{r^{3}}, \ \ \ \ \ \
 r,\,F,\, G > 0\,.
 \label{unsolvable}
 \ee
At a large value of $G \gg 1$, potential $W(r)$ may be represented by
its Taylor series again. Near its absolute minimum which occurs at
$r_{min}={R}=(G/F)^{1/6} \gg 1$ we have
  \be
 W({{R}}+\xi)=2\,\sqrt{G\,F} +
 9\,\sqrt{G\,F}\,\frac{\xi^2}{{R}^2}  -
 9\,\sqrt{G\,F}\,\frac{\xi^3}{{R}^3}
 +\ldots\,.
 \label{jedna}
 \ee
This expansion illustrates the feasibility of the same
harmonic-oscillator approximation approach to our generic
``unsolvable example" (UE). In the light of the large$-\nu$
equivalence SHO $\leftrightarrow$ RHO between the two exceptional
but ``equally exactly" solvable examples, the unsolvable equation
(\ref{unsolvable}) and its asymptotic solvability opens a way
towards a triple asymptotic equivalence between the {\em three}
low lying spectra,
 \be
 {\rm UE}\ \
 \longleftrightarrow\ \
 {\rm RHO}\ \
 \longleftrightarrow\ \
 {\rm SHO}\,.
 \label{tripleq}
 \ee
This means that we may, alternatively, approximate also any
unsolvable model by the zero-order solvable model with a singularity.
Of course, the triple-osculation property (\ref{tripleq}) restricts
our choice of the spiked SHO model characterized, for clarity, by the
zero subscript~$_0$,
  \be
 W_0(R_0+\xi)=
F_0\,(R_0+\xi)^{2}+\frac{G_0}{(R_0+\xi)^{2}}=
 2\,\sqrt{G_0\,F_0} +
 4\,F_0\,\xi^2-4\,\sqrt[4]{\frac{F_0^5}{G_0}}\,\xi^3+\ldots\,.
 \label{nula}
 \ee
We must require that the minima are the same, $R_0 = {{R}}$, and that
the two harmonic oscillator parts of the wells (\ref{jedna}) and
(\ref{nula}) differ at most by a convenient constant shift. These
requirements form a set of two equations with easy solution,
  \ben
 F_0=\frac{9}{4}\,\sqrt[6]{G\,F^5}, \ \ \ \ \ \ \ \
 G_0=\frac{4}{9}\,\sqrt[6]{G^5\,F}\,.
 \een
In the domain of the large $G\gg 1$ we have $G_0 \gg F_0$ as
expected.

We may summarize that all the unsolvable OD examples may be studied
via {\em different} implementations of the {\em same} idea of the
strong-repulsion approximation. Of course, even when all the
available free parameters coincide as they should, one may still test
and compare the {\em practical numerical} performance of the RHO and
SHO alternatives. This will not be pursued here. We only shortly
emphasize that the use of the more sophisticated two-parametric SHO
zero-order forces with a core gives a better chance for an optimal
choice of the zero-order approximant. {\it A priori}, an additional
benefit might be spotted in the better coincidence of the domains
(i.e., the half-axes of the coordinates) which might also support the
preference of the pair of the UE and SHO Hamiltonian operators. Thus,
the similarity of the shape (often called osculation) of the given
curve $W^{(UE)}$ near its minimum with the solvable spiked well
$W^{(SHO)}_0$ appears to be better founded, in spite of its slightly
less immediate construction. Presumably, the importance of the
similar arguments might further increase at $A=3$, in the genuine PD
context.

\section{Three-body spectra at any $K$ and $L$}

\subsection{Solvable guide: The $K=L=2$ model at
$\nu \gg 1$ }

The only non-constant angular dependence in the Calogero's
$\Omega^{(SHO)}(\r,\a)$ occurs in its short-range repulsion
component $\Omega^{(-2)} (\r,\a)\sim \sin^{-2} 3\a$. For $\a \in
(0, \pi/3)$ (i.e., within a physical wedge in the phase space
which is called, sometimes, the Weyl's chamber), this function has
a unique minimum at $\a_{min}=\pi/6$ where $\sin 3\a_{min} = 1$.
In its vicinity, the angular Taylor series may be calculated,
 \ben
 \frac{1}{\sin^2 3(\gamma +\pi/6)} =1 +9\,\gamma^2+ 54\,\gamma^4
 + \frac{1377}{5}
 \,\gamma^6 + \ldots\,.
 \een
This means that the absolute minimum of the potential
$\Omega^{(SHO)}(\r,\a)$ will lie at $\a=\a_{min}=\pi/6$. For the
determination of its second coordinate $\r= {R}$, we may analyze the
function $ \Omega^{(SHO)}(\r,\pi/6) =V_{eff}(\r)$ of the single
variable $\r>0$. The second necessary Taylor series is then very
easily derived,
 \ben
 V_{eff}(\r)=
 \frac{9\nu(\nu+1)}{2\r^2} + 3\,\omega^2\r^2
 = V_{eff}({R})+ \frac{1}{2} (\r-{R})^2\,
 V''_{eff}({R}) + {\cal O}\left [(\r-{R})^3\, V'''_{eff}({R})\right ]
 \een
 \be
= 6\omega^2{R}^2+12 \omega^2(\r-{R})^2 + {\cal O}\left
 [\frac{(\r-{R})^3}{{R}}\right ],\ \ \ \ \ \ \ \ \ \ \ \
 {R} = \sqrt[4]{
 \frac{3\nu(\nu+1)}{2\omega^2} }\,.
 \ee
Obviously, we just have to assume that ${R} \gg 1$ in order to
achieve a full consistency of the Calogerian model with its
two-dimensional RHO {\em separable} approximation,
 \be
  \Omega^{(SHO)}({R}+\xi,\eta/{R} +\pi/6)
 \approx
 6\omega^2{R}^2+12 \omega^2\xi^2 + 27\,\omega^2\,\eta^2
 \label{rescaled}
 \,.
 \ee
In a way predicted by the single-particle guide of section 2 one
discovers that all the higher-order terms ${\cal O}( \eta^m \times
\xi^n)$ remain suppressed by the same factor $1/R^{m+n-2} \ll 1$.
Thus, in the domain of large $\nu$, the low-lying spectrum of
energies may be approximated by the RHO formula
 \be
 E=E_{m,n}=6\omega^2R^2+2\sqrt{3}\,\omega\,(2n+1) +
 3\sqrt{3}\,\omega(2m+1) + \ldots\,.
 \ee
Up to the higher-order corrections, this prediction ``explains" the
well known exact spectrum of the low-lying energies in the Calogero
model  \cite{Calogero}.

We may conclude that in spite of the use of the polar coordinates,
the model is suitable for the two-dimensional RHO approximation using
the original cartesian coordinates rotated merely by $\pi/6$. This is
an immediate consequence of the structure of the kinetic term in eq.
(\ref{PDEpol}),
 \ben
  -\frac{\p^2}{\p \r ^2} -
 \frac{1}{\r ^2}
 \frac{\p^2}{\p \a ^2}\,.
 \een
Obviously, it may be {\em locally} re-interpreted as a cartesian
kinetic energy,
 \ben
  -\frac{\p^2}{\p \r ^2} -
 \frac{1}{(R+\xi)^2}
 \frac{\p^2}{\p \a ^2}=
  -\frac{\p^2}{\p \xi ^2} -
  \left [
  1+{\cal O}
  \left (
  \frac{\xi}{R}
  \right )
  \right ]
 \frac{\p^2}{\p \eta^2}
 \een
with the re-scaling of one of the coordinates which we employed also
in eq.~(\ref{rescaled}) above.

\subsection{$A=3$ confinement and its sample proof
for the non-separable $L=3$ oscillator}

One of the simplest non-separable three-body models may be based
on the cubic anharmonic-oscillator two-body potentials
 \be
 V^{(CAHO)}(x_i-x_j) =
  \omega^2\, \left( x_i-x_j \right )^2
 +
 \gamma\,\, \left( x_i-x_j \right )^3
 +
  \frac{\nu(\nu+1)}{\left( x_i-x_j \right )^2}\,.
  \label{star}
 \ee
Of course, one must proceed with a certain care since even in the
single-particle case, the cubic potential is not confining on the
real line \cite{Alvarez}. In the present setting, fortunately, we
need not work on the whole real line. On the contrary, due to the
presence of the strong repulsion,  we {\em must} parallel the
Calogero's considerations \cite{Calogero} and fix the ordering {\em
or} numbering of our three particles in advance, keeping in mind that
in quantum case, our particles cannot tunnel through the
$1/(x_i-x_j)^2$ barriers. This means that we may demand that, say,
$x_1 > x_2 > x_3$. This automatically guarantees that our system will
``live" in a single ``physical" wedge with $\a \in (0,\pi/3)$. At
this moment we may also guarantee that the cubic component
$\Omega^{(3)}(\r,\a)$ of the whole potential $\Omega^{(CAHO)}(\r,\a)$
will remain non-negative (i.e., confining or vanishing) at all the
positive couplings $\gamma > 0$.

In a certain challenging extreme, let us now drop the safely
confining quadratic force in eq. (\ref{star}) and show that even at
$\omega=0$, the three-body bound states will still remain localized.
In essence, we have to demonstrate that the spiked cubic (SC)
two-body forces $V^{(SC)}(x_i-x_j)$ [$=V^{(CAHO)} (x_i-x_j)$ at
$\omega=0$] will guarantee the confinement of the three-body system
even at the ends of the segment $\a \in (0,\pi/3)$ where the
potential $\Omega^{(3)}(\r,\a)$ itself vanishes.

The proof proceeds as follows. Firstly, assuming that $\r$ is
sufficiently large we show that with respect to the angle $\a$, there
exist two minima {in the {\em complete} SC potential}
 \be
 \Omega^{(SC)} (\r,\a)
 = \alpha^2\,\r^3 \sin 3\a + \frac{\beta^2}{\r^2\sin^2 3\a},
 \ \ \ \ \ \ \ \ \alpha^2=\frac{3}{2}\,\gamma > 0, \ \ \
 \beta^2= \frac{9}{2}\nu(\nu+1) > 0\,.
 \ee
They occur at the two $\r-$dependent angles $\a_\pm=\pi/6-\a_\mp$.
Their $\r-$dependence is controlled by the rule $\p_\a
\Omega^{(SC)}=0$ with the following solution and its consequence,
 \ben
 \sin^2 3 \a_\pm = \frac{2\beta^2}{\alpha^2\r^5}
 \ \ \
 \Longrightarrow
 \ \ \
 \Omega^{(SC)} > \frac{3\r}{2}\left (2\alpha^4\beta^2\r\right )^{1/3},
 \ \ \ \ \r \geq \r_0=
 \left ( \frac{2\beta^2}{\alpha^2} \right )^{1/5}\,.
 \een
This means that the potential $\Omega^{(SC)}$ grows sufficiently
quickly as a function of $\r$ in the  whole $\r \gg 1$ asymptotic
wedge. The model possesses just the discrete spectrum.

Marginally we may note that the latter proof may be generalized from
the above very special cubic model to all the potentials
$\Omega^{(2M+1)}(\r,\a)$. Indeed, as long as the angular minimum
moves quite quickly to the wedge boundary with the growing $M$, $
\a_- =\pi/3-\a_+ \sim \r^{-(2M+3)/3}$, the insertion of this value in
any $M > 1$ potential produces the general estimate
 \ben
 \Omega = \alpha^2\,\Omega^{(2M+1)} + \beta^2\,
  \Omega^{(-2)} \sim \r^{4M/3}, \ \ \ \ \ \ \ \
 \ \ \ \r \gg 1\,.
 \een
We  see that for all the integer exponents $M> 0$, no particle can
escape in infinity. The potential grows at least as a $4M/3-$th power
of $\r$ in the whole interval of the angles in the asymptotic domain
of $\r$.

\subsection{An ultimate illustration: $A=3$
 spectrum at $K=L=3$}

The climax of our analysis and considerations comes with the
strong-repulsion construction of the energies in any power-law $A=3$
system. For the sake of brevity we shall pick up, {\it pars pro
toto}, the same unsolvable spiked cubic oscillator as above,
described by the non-separable PD Schr\"{o}dinger equation
 \be
 \left [
  -\frac{\p^2}{\p \r ^2} -
 \frac{1}{\r ^2}
 \frac{\p^2}{\p \a ^2}+
 \Omega^{(SC)} (\r,\a)
 \right ]\,\psi(\r,\a) =
 E\,\psi(\r,\a),
 \ \ \ \ \ \ \beta^2 \gg 1\,.
 \label{SCE}
 \ee
We already know that its potential has a double-well shape with two
minima at every fixed and sufficiently large radial distance $\r\gg
1$ from the origin. With the decrease of $\r$, both these minima
decrease and their positions move towards the central maximum (which
``sits" at $\a = \a_0=\pi/6$). All these three extremes merge at
$\r=\r_0= \left ( {2\beta^2}/{\alpha^2} \right )^{1/5}$. Below this
distance, just a single minimum survives in the middle of the valley,
at $\a=\a_0$.

The potential at its minimum still decreases with the further
decrease of $\r$, and the unique absolute minimum of $ \Omega^{(SC)}
(\r,\a)$ is finally reached at $\a=\a_0=\pi/6$ and at the distance
 \ben
 R=\r^{(min)}_0= \sqrt[5]{2\beta^2/(3\alpha^2)}.
 \een
This quantity is large whenever $\beta^2 \gg 1$ so that we may use it
as a new, slightly simpler measure of  the strength of the SC
two-body repulsion. This means that we may re-define $\beta^2 =
3\,\alpha^2\,R^5/2$, $\r=R+\xi$, $\a=\pi/6+\eta/R$ and expand
 \ben
 \frac{1}{\alpha^2}\,
 \Omega^{(SCO)} (\r,\a)
= \r^3 \sin 3\a + \frac{\beta^2}{\alpha^2\,\r^2\sin^2 3\a}
=
 \een
 \ben
 =\frac{5}{2}
 {R}^{3}+9\,R{{ \eta}}^{2}+\frac{15}{2}\,R{\xi}^{2}-
 \een
 \ben
 -{\frac
 {81}{2}}\, \xi\,{{ \eta}}^{2}-5\,{\xi}^{3}+
 \frac{1}{R}\,
 \left (
 {\frac {675}{8}}\,{{{}{{ \eta} }^{4}}}+27\,{{{}{\xi}^{2}{{
 \eta}}^{2}}}+\frac{15}{2}\,{ {{}{ \xi}^{4}}}
 \right )
 +{\cal O}
 \left (
 \frac{1}{R^2}
 \right )\,.
 \een
We see that the corrections decrease with the integer powers of
$1/R$. This enables us to insert their series in eq. (\ref{SCE}).
With a new re-scaling constant $\sigma = 1/\sqrt[4]{R}$ introduced in
the two independent variables,
 \ben
 \r = R + \sigma\,x, \ \ \ \ \ \ \ \ \a=\pi/6+\sigma\,y/R\,,
 \een
this gives our final approximate Schr\"{o}dinger equation
 \be
 \left [
  -\frac{\p^2}{\p x^2} -
 \frac{\p^2}{\p y^2}+
 \alpha^2
 \left (
 \frac{15}{2}\,{x}^{2}
 +9\,{{y}}^{2}
 \right )
 +{\cal O}
 \left (
 \sigma^5
 \right )
 \right ]\,\psi(x,y) =
 \sigma^2\,\left (
 E-\frac{5}{2}
 {R}^{3}
 \right )
 \,\psi(x,y)\,.
 \label{SCEap}
 \ee
The reliability of this approximation will increase with the growth
of $R \gg 1$ or $\sigma=\sigma(R) \ll 1$.

At this point one must emphasize the {\em two} parallel merits of the
approximation (\ref{SCEap}). Firstly, its is a separable PD equation
which may be analyzed as two independent OD equations. Secondly, both
these OD components are exactly solvable. This means that in the
strongly spiked domain, the low-lying energies are given by the
following closed formula
 \be
 E=E_{m,n}= \frac{5}{2}\,R^3+
 \alpha\,\sqrt{\frac{15\,R}{2}}\,(2m+1) + 3\,\alpha\,\sqrt{R}\,(2n+1)
 + {\cal O}
 \left (
 \frac{1}{R^{3/4}}
 \right )\,.
 \label{energies}
 \ee
We see that these energies are numbered by the respective ``radial"
and ``angular" quantum numbers $m,n=0,1,\ldots $ in a way which
resembles the previous Calogerian exactly solvable case. Thus, one
may expect that the same asymptotic construction remains applicable
to all our models (\ref{PDE}).

\section{Summary}

By our present paper, two main messages are delivered:

\begin{itemize}

\item
the power-law confinement of three spinless particles in one
dimension leads to a comparatively simple PD Schr\"{o}dinger
equation, provided only that we work in the spherical Jacobi
coordinates;

\item
a quantum osculation method generates the low-lying spectra with the
precision which increases with the strength of the repulsive two-body
core in $V(x_i-x_j)$.

\end{itemize}

 \noindent
An exceptional, unexpected separability of the quartic anharmonic
oscillations of three particles has been also revealed as a byproduct
of our systematic explicit construction of the corresponding PD
Hamiltonian operators.

The former pair of messages resulted, basically, from a successful
generalization of the Calogero's trigonometric identity
(\ref{keyr}). In the historical perspective of ref.
\cite{Calogero}, the latter identity opened the path towards the
explicit construction of the whole class of solvable models. Some
of the consequences of its present generalizations are equally
exciting: This is the essence of our mathematical achievement.

One of its most important uses for physics may be seen in the
context of the large$-\beta$ expansions. Their construction  at
any $K \neq 2 \neq L$ seems to be significantly facilitated by a
generic survival of the separability {\em beyond} the Calogero's
exceptional model whenever we move to the {\em  strongly repulsive
regime}. In this sense, even all the multi-term generalizations of
our present power-law $A=3$ oscillators characterized, in both the
long-range and short-range regime, by a polynomial behaviour of
their two-body potentials remain tractable as models which will be
{\em approximately separable and solvable} in the domain of the
very strong two-body repulsion at short distances.

One of the most remarkable purely technical aspects of our present
work appeared to be an {\em unexpectedly slow} growth of the
complexity of the closed trigonometric formulae for the total
potentials $\Omega^{(m)}(\r,\a)$ with the increase of absolute
values of their maximal-power superscripts $m$. This is a lucky
circumstance which enhances the feasibility of the construction of
the approximate spectra in polar coordinates and in the
strongly-spiked limit significantly.

In the future, one may expect that also a systematic incorporation
of the higher-order corrections will remain feasible. We plan the
more systematic study of such a possibility based on a refined
strong-coupling perturbation-series expansion of the observable
quantities in the powers of the auxiliary quantity $1/R$ where $R$
denotes the radial distance of the absolute minimum of
$\Omega^{(m)}(\r,\a)$ and grows in proportion to the strength of
the repulsive two-body core.

\subsection*{Acknowledgements}

Work supported by the grant Nr. A 1048302 of GA AS CR.

\newpage

\section*{Appendix A: Jacobi coordinates at $A = 3$}

It is well known that the introduction of the suitably normalized
centre-of-mass coordinate $Z=Z^{(A)} =\sum_{k=1}^{A} \, x_k /
{\sqrt{A}} $ enables the elimination of the free bulk motion. Of
course, the use of the new variable $Z$ imposes a constraint upon
the old $A-$plet of coordinates. The redundancy of one of the
positions $x_k$ may be easily resolved by the transition to the
multiplet $(Z,X,Y,\ldots)$ of the so called Jacobi coordinates
which may be defined at any number of particles $A$ and which we
shall specify, at $A=3$, by their most transparent matrix
definition
 \be
  \left (
 \ba Z\\X\\Y \ea
 \right )=
\left ( \begin {array}{ccc} 1/\sqrt {3}&1/\sqrt {3}&1/\sqrt {3 }\\
\noalign{\medskip}1/\sqrt {2}&-1/\sqrt {2}&0
\\
\noalign{\medskip}1/\sqrt {6}&1/\sqrt {6}&-2/\sqrt {6}
\end {array}\right )
 \left (
 \ba x_1\\x_2\\x_3 \ea
 \right )\
 \ee
with an elementary inversion. In this manner, the original
particle coordinates may be understood as a triplet of linear
functions of three new parameters,
  \be
 \left (
 \ba x_1(Z,X,Y)\\
\noalign{\medskip}x_2(Z,X,Y)\\
\noalign{\medskip}x_3(Z,X,Y) \ea
 \right )
= \left ( \begin {array}{ccc} 1/\sqrt {3}& 1/\sqrt {2}& 1/\sqrt
{6}
\\
\noalign{\medskip} 1/\sqrt {3}& -1/\sqrt {2}& 1/\sqrt {6}
\\
\noalign{\medskip} 1/\sqrt {3}& 0& -2/\sqrt {6}
\end {array}\right )
 \left (
 \ba Z\\
\noalign{\medskip}X\\
\noalign{\medskip}Y \ea
 \right )\,.
 \ee
These functions enter the general interaction term in eq.
(\ref{SE}),
 \ben
\hat{W}(x_1,x_2, x_3)=
 \sum_{m=-K}^{L}\,F_m\,W^{(m)}(x_1,x_2, x_3)\,,
 \ \ \ \ \ \ \ \ F_L=\alpha^2> 0 , \ \ \
 F_{-K}=\beta^2 > 0,
  \een
 \be
 W^{(m)}
(x_1,x_2,x_3)= (x_1-x_2)^m
 +(x_2-x_3)^m
 + (x_3-x_1)^m\,
  \label{anhCalogero}
 \ee
where the positivity of $F_L$ guarantees the confinement while the
positivity of $F_{-K}$ prevents the system from a collapse. In
such a notation we may re-write our $A=3$ Schr\"{o}dinger equation
as a reduced PD problem in the two cartesian coordinates $X$ and
$Y$,
  \be
\left \{ -\frac{\p^2}{\p X^2} -\frac{\p^2}{\p Y^2} +
  \sum_{m=-K}^{L}\,
 F_m\,   {U}^{(m)}(X,Y)
 -E \right \}
\,\Phi(X,Y)=0\,
 \label{PDE}
 \ee
where the functions
  \ben
 {U}^{(m)}(X,Y) \equiv
  W^{(m)}
  \left [ x_1(Z,X,Y), x_2(Z,X,Y), x_3(Z,X,Y) \right ]\,
 \een
prove independent of $Z$. One may also employ the polar
re-parametrization of the Jacobi coordinates,
 \be
 X=X(\r,\a)=\r  \, \sin \a, \ \ \ \ \ \ Y =Y(\r,\a)=
  \r  \,  \cos \a\,,
 \label{polar}
 \ee
arriving at an alternative, equivalent formulation of the same
Schr\"{o}dinger equation,
  \be
\left \{ -\frac{\p^2}{\p \r ^2} -
 \frac{1}{\r ^2}
 \frac{\p^2}{\p \a ^2} +
  \sum_{m=-K}^{L}\,
 F_m\,  {\Omega}^{(m)}(\r,\a)
 -E \right \}
\,\Phi[X(\r,\a),Y(\r,\a)]=0\,
 \label{PDEpol}
 \ee
where we introduced another abbreviation
 \ben
{\Omega}^{(m)}(\r,\a) \equiv {U}^{(m)} [X(\r,\a), Y(\r,\a)] \,.
 \een

\section*{Appendix B. Auxiliary trigonometric identities}

Any practical analysis of the Schr\"{o}dinger $A=3$ equation
(\ref{SE}) in Jacobi coordinates requires the explicit
specification of its interaction term (\ref{tribal}). Separately,
let us derive its form for the three specific subclasses of the
underlying two-body input force.

\subsection*{B.1. Potentials ${U}^{(m)}$ and $\Omega^{(m)}$ with even $m=2M$
}

In our present notation, the regular two-body interactions are
characterized by the even integer superscripts and by the
following asymptotically confining interaction functions,
 \ben
 {U}^{(2M)}(X,Y)
 = \left [x_{{1}}(R,X,Y)-x_{{2}}(R,X,Y) \right ]^{2M}+
 \een
 \ben
 +
 \left [x_{{2}}(R,X,Y)-x_{{3}}(R,X,Y) \right ]^{2M}
 +
 \left [x_{{3}}(R,X,Y)-x_{{1}}(R,X,Y) \right ]^{2M}\,.
 \een
The growth of $M$ means a strengthening of the confinement in both
infinities, which has a definite phenomenological appeal. The
explicit form of the functions $ {U}^{(2M)}(X,Y)$ is,
unfortunately, more and more complicated for the larger and larger
$M$,
 \[
{U}^{(6)}(X,Y)={\frac {33}{4}}\,{X}^{6}+{\frac
{45}{4}}\,{X}^{4}{Y}^{2}+{\frac {135}{ 4}}\,{X}^{2}{Y}^{4}+{\frac
{27}{4}}\,{Y}^{6},
\]
 \[
{U}^{(8)}(X,Y)=\frac{3}{8}\,\left ({X}^{2}+{Y}^{2}\right )\left
(27\,{Y}^{6}+225\,{X}^{2}{Y}^
{4}-15\,{X}^{4}{Y}^{2}+43\,{X}^{6}\right ) ,
\]
etc. Still, our search for the parallels between the solvable and
unsolvable models finds its reward here since with the polar
coordinates where we introduce an abbreviation
 \ben
 {\Omega}^{(2M)}(\r,\a)=
 {U}^{(2M)}[X(\r,\a),Y(\r,\a)]
 =
 \een
 \ben =\left \{x_{{1}}\left [R,X(\r,\a),Y(\r,\a)\right ]-x_{{2}}\left [
 R,X(\r,\a),Y(\r,\a)\right ] \right \}^{2M}+
 \ldots
 \een
the fairly perceivable simplifications  emerge after the patient
trigonometric manipulations. This is one of our most important
technical results. We get the sequence of the pleasantly compact
formulae,
 \[
{\Omega}^{(6)}(\r,\a)= \frac{3\,\r ^6}{4} \left (9+2\,\sin^2 3\a
\right ) = \frac{3\,\r  ^6}{4} (10-\cos\,6\,\a) ,
\]
 \[
{\Omega}^{(8)}(\r,\a)=\frac{3\, \r  ^8}{8}\,\left ( 27+16\,\sin^2
 3\a \right
) \, ,
\]
 \[
{\Omega}^{(10)}(\r,\a)={\frac {27\,\r  ^{10}}{16}}\, \left ( 9
 + 10\, \sin^2
3\a \right ) \,,
\]
 \[
{\Omega}^{(12)}(\r,\a)={\frac {729\,\r  ^{12}}{32}}+{\frac {81\,\r
^{12}}{2}}\,\sin^2 3\,\a +\frac{3\,\r  ^{12}}{4}\,\sin^4 3\,\a\,,
\]
etc. A remarkable regularity emerges in this pattern. For example,
the choice of the superscripts $m = 2M=6N$, $6N+2$ or $6N+4$ leads
to the compact general formula
 \be
 \frac{{\Omega}^{(2M)}(\r,\a)}{\r^{2M}}
 = \frac{3^M}{2^{M-1}}+ c_1\ \sin^{2}
 3\a +\ldots +
c_N\, \sin^{2N} 3\a \,
 \ee
with a $\r-$independence on the right hand side and with an
extremely elementary $N-$ and $M-$dependence of some its
coefficients which may be derived very easily whenever needed.

\subsection*{B.2. Potentials $U^{(m)}$ and $\Omega^{(m)}$ with odd
$m=2M+1$ }

Systematic study of all the potentials $U^{(m)}$ with the {\em
odd} powers $m= 3, 5, \ldots$ is not so easily extended to $A=4$
but in the present $A=3$ setting it is well defined and worth the
study. The elementary experimenting teaches us quickly that up to
the trivial ${U}^{(1)}(X,Y)=0$ we have to deal with the
interesting functions. One cannot be surprised by the observation
that the complexity of the representation of the complete
potentials in Jacobi coordinates grows quite quickly with their
superscript,
 \[
{U}^{(3)}(X,Y)=\frac{3}{2}\,\sqrt {2}\,X\left
({X}^{2}-3\,{Y}^{2}\right )\, ,
\]
 \[
{U}^{(5)}(X,Y)={\frac {15}{4}}\,\sqrt {2}\left
({X}^{2}-3\,{Y}^{2}\right )\left ({X}^ {2}+{Y}^{2}\right )X ,
\]
 \[
{U}^{(7)}(X,Y)={\frac {63}{8}}\,\sqrt {2}\left
({X}^{2}-3\,{Y}^{2}\right )X\left ({X} ^{2}+{Y}^{2}\right )^{2} ,
\]
 \[
{U}^{(9)}(X,Y)=\frac {3}{16}\,\sqrt {2}\left
(81\,{Y}^{6}+279\,{X}^{2}{Y}^{4}+219\,{X}^{4}{Y}^
{2}+85\,{X}^{6}\right )\left ({X}^{2}-3\,{Y}^{2}\right )X
\]
etc. Fortunately, in the same manner as above, the use of the
polar coordinates makes the geometric structure and symmetries of
these two-dimensional potential wells much more evident,
 \[
 {\Omega}^{(3)}(\r,\a)=
  \frac{3\,\r  ^{3}}{2}\,\sqrt {2}\,\sin 3\a ,
\]
 \[
{\Omega}^{(5)}(\r,\a)= {\frac {15\,\r  ^{5}}{4}}\,
 \sqrt {2}\,\sin 3\a ,
\]
 \[
{\Omega}^{(7)}(\r,\a)= {\frac {63\,\r  ^{7}}{8}}\,\sqrt {2}\,
 \sin 3\a ,
\]
 \[
{\Omega}^{(9)}(\r,\a)= \frac{3\,\r  ^9}{16}\,\sqrt {2}\,\sin 3\a
 \,\left (
81+4\,\sin^2 3\a \right ) \
 \left [
=\frac{3\,\r  ^9}{16}\,\sqrt {2}\,\sin
 3\a
\,\left ( 83-2\,\cos\,6\,\a \right ) \right ],
\]
 \[
{\Omega}^{(11)}(\r,\a)={\frac {33\,\r  ^{11}}{32}}\,\sqrt
{2}\,\sin 3\a \, \left ( 27+4\,\sin^2 3\a \right )\,,
\]
 \[
{\Omega}^{(13)}(\r,\a)= {\frac {117\,\r  ^{13}}{64}}\,\sqrt
 {2}\,\sin 3\a \,
\left (27+8\,\sin^2 3\a \right )\,,
\]
etc. Again, once we have $m = 6N+3$ or $m=6N+5$ or $m =6N+7$, the
explicit form of the potential ${\Omega}^{(m)}$ will be prescribed
by the trigonometric formula
  \be
  \frac{{\Omega}^{(2M+1)}(\r,\a)}{\r^{2M+1}}
  = (2M+1)\,\sqrt{2}\,\sin \,3\a\, \left
(\frac{3^{M-1}}{2^{M}}+ d_1\ \sin^{2} 3\a +\ldots + d_N\,
\sin^{2N} 3\a \right ) \ee where, formally,
  \ben
  \ \ \ \ \ \ \  N = entier \left [ \frac{M-1}{3} \right ]\,,
  \ \ \ \ \ \ \ \ \ \ \
M = 1, 2, \ldots\,
 \een
and $entier[x]$ denotes the integer part of a real number $x$.

\subsection*{B.3. Strongly
singular potentials $\Omega^{(m)}$ with negative $m$}

Using the same abbreviations as above we may immediately
complement the Calogero's identity (\ref{singular}) by its
Coulombic predecessor,
 \[
{\Omega}^{(-1)}(\r,\a)= - \frac{3}{\r  \,\sqrt {2}\,\sin 3\a } ,
\]
 \be
{\Omega}^{(-2)}(\r,\a)= \left (\frac{3}{\r  \,\sqrt {2}\,\sin 3\a
}\right )^{2}=\left [ {\Omega}^{(-1)}(\r,\a) \right ]^2.
 \label{keyr}
 \ee
These two formulae exhaust the set of the standard singular
forces. Nevertheless, there is no physical reason for avoiding the
more singular repulsion near the origin, and the simplicity of the
latter two formulae attracts attention to the more general spikes
in
 \be
 {\Omega}^{(-k)}(\r,\a)= \left (x_{{1}}-x_{{2}} \right )^{-k}+
 \left (x_{{2}}-x_{{3}} \right )^{-k}+
\left (x_{{3}}-x_{{1}} \right )^{-k} \,.
 \ee
At the first three ``nontrivial"  exponents $k > 2$ we reveal
again the same overall structure of the interaction terms in their
trigonometric representation,
 \[
{\Omega}^{(-3)}(\r,\a)= \left [ {\Omega}^{(-1)}(\r,\a) \right ]^3
\left ( 1 - \frac{4}{9}\, \sin^2\,3\,\a \right )
 ,
\]
 \[
{\Omega}^{(-4)}(\r,\a)= \left [ {\Omega}^{(-1)}(\r,\a) \right ]^4
\left ( 1 - \frac{16}{27}\, \sin^2\,3\,\a \right )
 ,
\]
 \[
{\Omega}^{(-5)}(\r,\a)= \left [ {\Omega}^{(-1)}(\r,\a) \right ]^5
\left ( 1 - \frac{20}{27}\, \sin^2\,3\,\a \right )
 ,
\]
Obviously, the pattern remains precisely the same as above. In
place of any routine classifications, let us only add the first
less trivial formula
 \[
{\Omega}^{(-6)}(\r,\a)= \left [ {\Omega}^{(-1)}(\r,\a) \right ]^6
\left ( 1-{\frac {904}{667}}\,\sin^{2} \,3\,\a +{\frac
{8248}{18009}}\,{\sin}^{4}\,3\,\a \right )
 ,
 \]
which nicely illustrates not only the emergence of the new terms
but also the close parallel between the positive and negative
superscripts.

\end{document}